\documentclass[%
 reprint,
superscriptaddress,
nofootinbib,
 amsmath,amssymb,
 aps,
 prb,
]{revtex4-2}

\usepackage{makecell} 
\usepackage[colorinlistoftodos]{todonotes} 
\usepackage{natbib}

\usepackage{subfiles}
\usepackage{float}
\usepackage{graphicx}
\usepackage{dcolumn}
\usepackage{bm}
\usepackage{multirow}
\usepackage{lipsum} 

\usepackage{amsmath}
\let\oldAA\AA
\renewcommand{\AA}{\text{\normalfont\oldAA}}

\newcommand{\bx}{\bold{x}}
\newcommand{\br}{\bold{r}}
\newcommand{\bR}{\bold{R}}
\newcommand{\bom}{\bold{\Omega}}

\begin{document}


\title{Fermionic Neural Network with Effective Core Potential}

\author{Xiang Li}
\affiliation{ByteDance Inc, Zhonghang Plaza, No. 43,  North 3rd Ring West Road, Haidian District, Beijing.}
\email{\{lixiang.62770689, renweiluo\}@bytedance.com}
\author{Cunwei Fan}
\affiliation{Department of Physics and Institute for Condensed Matter Theory, University of Illinois 1110 W. Green Street, Urbana, IL 61801, U.S.A.}
\email{cfan11@illinois.edu}
\author{Weiluo Ren}
\affiliation{ByteDance Inc, Zhonghang Plaza, No. 43,  North 3rd Ring West Road, Haidian District, Beijing.}
\author{Ji Chen}
\affiliation{%
School of Physics, Peking University, Beijing 100871, People’s Republic of China
}
\email{ji.chen@pku.edu.cn}

\date{\today}
\begin{abstract}
Deep learning techniques have opened a new venue for electronic structure theory in recent years.
In contrast to traditional methods, deep neural networks provide much more expressive and flexible wave function ansatz, resulting in better accuracy and time scaling behavior. 
In order to study larger systems while retaining sufficient accuracy, we integrate a powerful neural-network based model (FermiNet) with the effective core potential method, which helps to reduce the complexity of the problem by replacing inner core electrons with additional semi-local potential terms in Hamiltonian. In this work, we calculate the ground state energy of 3d transition metal atoms and their monoxides which are quite challenging for original FermiNet work, and the results are in good consistency with both experimental data and other state-of-the-art computational methods. 
Our development is an important step for a broader application of deep learning in the electronic structure calculation of molecules and materials.
\end{abstract}

\maketitle

\section{introduction}
The last decade has witnessed incredible fast development of artificial intelligence. Deep learning \cite{Goodfellow-et-al-2016} becomes widely used and receives great success in computer vision \cite{10.1145/3065386,NIPS2015_14bfa6bb}, natural language processing \cite{NIPS2017_3f5ee243,6949403} and recommendation systems \cite{10.1145/2959100.2959190,gao2021deep}, just to name a few. In the past few years, deep learning technology is broadly applied in computational physics and chemistry to tackle key challenges in \textit{ab initio} molecule modelling \cite{han2017deep,schnet,deepmd,deepwf,deepbf,pfau2020ab,spencer2020better,Hermann2020,simple_trial,Choo2020FermionicNS,wilson2021simulations,deepks}, which are crucial for materials design, drug discovery and other applications.

 Deep learning techniques can be roughly divided into two categories. 
 The first category, e.g. machining learning force field, aims at improving the efficiency of simulation while retaining the accuracy at a higher level of theory \cite{schnet,han2017deep,deepmd}. 
 These methods usually requires labeled data (such as energy and force of a given structure), and then trains the neural network by minimizing the deviation between its prediction and the labeled data.
 The second category, e.g. neural network based quantum Monte Carlo (QMC) \cite{deepwf,deepbf,pfau2020ab,spencer2020better,Hermann2020,Choo2020FermionicNS,simple_trial}, targets more accurate electronic structure and only needs unlabeled sample data in training.
 A deep neural network can provide much more expressive and flexible wave function ansatz than traditional forms used in QMC, which leads to better accuracy. 
 Recently developed PauliNet and FermiNet are two promising examples \cite{pfau2020ab,spencer2020better,Hermann2020}, which have shown the ability to outperform traditional methods such as coupled cluster with single, double and perturbative triple excitations [CCSD(T)] in certain systems. 
 In spite of their advantages, these methods also have their own drawbacks: the enormous amount of parameters in deep neural networks strictly restrict the simulation speed and system size they can study. For example, it would take the FermiNet a month to simulate a system of about 30 electrons using 16 V100 GPUs in its TensorFlow version \cite{pfau2020ab}.

In order to extend neural network electronic structure calculations to larger systems, the computation complexity has to be reduced and one helpful approach is the so-called effective core potential (ECP) method (also known as pseudopotential). Electrons in each system can be divided into core electrons and valence electrons. Core electrons, filling inner shells of the system, are tightly bound around the atom cores, and it is mainly the valence electrons in the outer shell that determine the property of the system. The ECP method simply removes the core electrons from computation and introduces semi-local potential terms to effectively simulate their influence on valence electrons, in which way the number of electrons in the calculation is reduced and the whole computation process is accelerated. 
ECP method is widely employed in traditional electronic structure calculations, such as density functional theory (DFT), post Hartree-Fock and QMC.
In particular, the development of ECPs for QMC are still a hot subject of research. Nevertheless, there are already a number of ECPs designed or used in QMC calculations, such as Burkatzki-Filippi-Dolg (BFD) ECP \cite{bfdecp}, Stuttgart (STU) ECP \cite{stuecp}, Trail-Needs (TN) ECP \cite{trail}, and correlation consistent (cc) ECP \cite{ecp_3d,ecp_al}.
These ECPs are examined carefully in traditional QMC simulations, but their numerical stability and performances have not yet been examined in deep neural networks, which has a much complex structure to express wave functions. 

In this work, we investigate the implementation of ECP method into a deep neural network, namely FermiNet. The main aim is to improve the efficiency of deep neural network modelling, so as to increase the size of the system that we can handle with FermiNet. Based on our method, we calculate the ground state energy of 3d transition metal atoms and their monoxides. The results show satisfactory consistency with both experimental data and other accurate \textit{ab initio} methods. We also discuss some details of how to improve the training efficiency of the ECP based FermiNet. 

The remainder of this paper is organized as follows. In Sec.~\ref{sec:method}, we introduce briefly the theoretical framework of neural network, ECP method, the workflow and calculation details. 
In Sec.~\ref{sec:result}, we present our calculations and results. 
In Sec.~\ref{sec:discus}, we discuss more details on our calculation and training.
In Sec.~\ref{sec:conclu}, we give a summary and outlook.

\section{METHOD}
\label{sec:method}
\subsection{Theoretical Framework}
Solving Schr\"{o}dinger equation is always the main task of electronic structure calculation. Under Born-Oppenheimer approximation, atomic motion is frozen and the equation for electron wave function $\psi$ can be formulated as
\begin{gather}
\label{eq:sch}
    \hat{H}\psi(\bx_1,\bx_2,\cdots,\bx_N)=E\psi(\bx_1,\bx_2,\cdots,\bx_N), \\
\begin{aligned}
    \hat{H}=&\sum_i -\frac{1}{2}\Delta_i + \frac{1}{2}\sum_{i\neq j}\frac{1}{|\br_i-\br_j|} \nonumber\\
        &- \sum_{i,I}\frac{Z_I}{|\br_i-\bR_I|} + \frac{1}{2}\sum_{I\neq J}\frac{Z_I Z_J}{|\bR_I-\bR_J|},
\end{aligned} 
\end{gather}
where $\bx_i=(\br_i,\sigma_i)$ denotes the spatial position and spin of i-th electron. $\bR_I,Z_I$ are the spatial position and charge of I-th nucleus, which are treated as external parameters for a given molecule structure. Moreover, since electrons obey Fermi-Dirac statistics, the wave function $\psi$ needs to be anti-symmetric with respect to the permutation of $(\bx_1,\bx_2,\cdots,\bx_N)$. The anti-symmetric requirement together with its high-dimensional nature make Schr\"{o}dinger equation notoriously hard to solve. In order to obtain highly accurate results, the time complexity of the state-of-the-art methods, such as CCSD, scales as large as $N^6$ or more, where $N$ is the total number of electrons in the system. 

Recently, deep neural network models, such as FermiNet and PauliNet \cite{pfau2020ab, spencer2020better, Hermann2020} are proposed and shed new light on the electronic structure problem. Based on the variational principle, these deep learning methods approach the ground state wave function via minimizing the energy expectation value $E_\theta$, which reads
\begin{gather}
\label{eq:energy}
    E_\theta = \frac{\int d^3\br\ \psi^*_\theta(\br)H\psi_\theta(\br)}{\int d^3\br\ \psi^*_\theta(\br)\psi_\theta(\br)}.
\end{gather}
$\psi_\theta$ is simply the wave function output by neural network and $\theta$ denotes all the parameters within. 

Traditional VMC approaches usually start from the Hartree-Fock wave function, which reads
\begin{gather}
\label{eq:hf}
    \psi_{\rm HF}(\bx)=\left|
    \begin{matrix}
    \phi_1(\bx_1)  & \cdots & \phi_1(\bx_N) \\
    \cdot & & \cdot\\
    \cdot & & \cdot\\
    \cdot & & \cdot\\
    \phi_N(\bx_1)  & \cdots & \phi_N(\bx_N) \\
    \end{matrix}
    \right|,
\end{gather} 
where $\{\phi_i\}$ denotes the molecular orbitals for the electrons. In deep neural networks \cite{deepbf,pfau2020ab,Hermann2020,spencer2020better}, the one electron orbital $\phi_i(\bx_j)$ is extended to $\phi_i(\bx_j;\bx_{\neq j})$, where $\bx_{\neq j}$ denotes all the electron coordinates except $\bx_j$, and it can be seen as a generalization of widely-used backflow transformation \cite{backflow}. In order to ensure the anti-symmetry of $\psi$, $\phi_i(\bx_j;\bx_{\neq j})$ is required to be permutation invariant with respect to permutation of ${\bx_{\neq j}}$.
With this generalization employed, the simulation accuracy is highly improved and the time complexity is retained as the traditional VMC approach, which scales as $N^4$.

However, existing deep neural networks, namely FermiNet and PauliNet, suffer from a large prefactor in its asymptotic $N^4$ time scaling. The enormous amount of parameters and linear operations within neural networks reduce their computation efficiency, and the largest system size they can study is limited up to dozens of electrons. In order to enlarge the system size we can study, it's natural to employ ECP method, which has already been widely used in quantum chemistry community. Within ECP framework, core electrons decouple with the valence electron, and additional semi-local potential terms are added in Hamiltonian to mimic core electron effects, which read
\begin{gather}
\label{eq:ecp}
    \hat{V}_{\rm ECP} = \sum_{v=1}^{n_v} V_{\rm loc}(r_v) + \sum_{v=1}^{n_v}\sum_{l=0}^{l_{\rm max}} V_l(r_v)\sum_{m=-l}^{m=l} |lm\rangle\langle lm|,
\end{gather}
where $n_v$ denotes the number of valence electrons and $|lm\rangle$ represents the spherical harmonics. Moreover, the potential terms $V_{\rm loc}$ and $V_{l}$ are simply functions of $r_v$, which represents the radial distance between valence electron and nucleus. These potential terms are usually expanded in Gaussian basis sets, which read


\begin{gather}
\label{eq:ecp_gau}
    V_l(r)=r^{-2}\sum_k A_{lk}r^{n_{lk}}e^{-B_{lk}r^2},
\end{gather}
where $A_{lk}, B_{lk}$ and $n_{lk}$ are the expansion parameters, and $l,k$ denote the angular quantum number and the expansion index, respectively. 

Note that effective core potentials usually diverge near nucleus due to the $r^{-2}$ term in Eq.~\eqref{eq:ecp_gau}, making the QMC simulation unstable. There are several kinds of ECPs such as ccECP, STU ECP, BFD ECP and TN ECP overcoming this problem via exact cancellation of diverging terms in Eq.~\eqref{eq:ecp}, which are more suitable for quantum Monte Carlo simulations \cite{stuecp,bfdecp,trail,ecp_al,ecp_3d}. In this work, we adopt recently proposed ccECP \cite{ecp_al,ecp_3d}. 
Other ECPs can be implemented in a similar way with FermiNet.

\subsection{Workflow}
The workflow of our work is an extension of FermiNet, with modifications related to the integrated ECP method.

The main workflow can be divided into three phases: pretrain, train and inference. In the pretrain phase, the neural network is trained to match the Hartree-Fock wave function, obtained via PySCF package \cite{pyscf} with certain ECP type specified. Then the neural network is trained to minimize the expected energy value. Note that with ECP employed, the gradient formula for energy optimization is slightly modified as follows,
\begin{equation}
\label{eq:grad}
\begin{gathered}
{\rm Grad} = \mathbb{E}_{\psi^2(\br)}[(E(\br)-\mathbb{E}_{\psi^2(\br)}[E(\br)])\nabla_\theta\log|\psi|],\\
E(\br) = \psi^{-1}(\br)\hat{H}\psi(\br)+E_{\rm nl}(\br),
\end{gathered}
\end{equation}
where $\mathbb{E}_{\psi^2(\br)}[\cdot]$ represents the expectation value according to the distribution $\psi^2(\bx)$ and it's evaluated via traditional Markov Chain Monte Carlo (MCMC) approach. Moreover, $E_{\rm nl}$ denotes the semi-local energy contribution from effective core potential and its specific form reads \cite{RevModPhys.73.33}
\begin{equation}
\label{eq:ecp_int}
\begin{aligned}
&\sum_{vlm}V_l(r_v)Y_{lm}(\bom_v)\int d\bom'_v Y_{lm}^*(\bom'_v)\frac{\psi(\br_1,\cdots,\br_v',\cdots,\br_N)}{\psi(\br_1,\cdots,\br_v,\cdots,\br_N)}.
\end{aligned}
\end{equation}
where $Y_{lm}$ denotes the spherical harmonics. This integral is over the solid angle $\bom_v'$ of valence electron vector $\br_v'$ with respect to the nucleus. 
Although, in the mean-field approximation, the integration has a closed-form result, an analytic result does not exist for the wavefunction $\psi$ produced by the FermiNet and thus numerical integration methods shall be used. After a long enough training process until the energy converges, the final energy result can be obtained from a separate inference phase, in which energy is estimated via pure MCMC approach without training.

\subsection{ECP Implementation Details}\label{ecpDetails}


%

Before carrying out the numerical integration concretely, we note that the integrand in Eq.~\eqref{eq:ecp_int} can be simplified further. The polar axis of the integrand can be set parallel to $\br_v$ and the outer $Y_{lm}(\bom_v)$ terms are reduced to constants \cite{fahy1990variational}, then terms in Eq.~\eqref{eq:ecp_int} read 
\begin{equation}
\label{eq:sim_int}
    \begin{aligned}
    \sum_m &V_l(r_v)Y_{lm}(\bom_v) \int d\bom'_v Y_{lm}^*(\bom'_v)\frac{\psi(\br_1,\cdots,\br_v',\cdots,\br_N)}{\psi(\br_1,\cdots,\br_v,\cdots,\br_N)} \\ 
    =~&V_l(r_v)\frac{2l+1}{4\pi}\int d\bom'_v P_l(\cos\theta'_v) \frac{\psi(\br_1,\cdots,\br_v',\cdots,\br_N)}{\psi(\br_1,\cdots,\br_v,\cdots,\br_N)},
    \end{aligned}
\end{equation}
where $P_l$ is the Legendre polynomial and $\theta'_v$ denotes the polar angle with polar axis set to $\br_v$ now. 
The integral in Eq.~\eqref{eq:sim_int} is over the unit sphere of $v$-th valence electron with respect to the nucleus and there are quite a few choices of integration quadrature suitable for this situation \cite{mitavs1991nonlocal,mclaren1963optimal,lebedev1975values,lebedev1976values}. We decided to use the 12-point icosahedron quadrature in the calculation because it has acceptable accuracy as well as a reasonably small number of points. In the 12-point icosahedron quadrature, 12 points and their weights are properly chosen so that the integration on a unit sphere is exact for integrands with only $l\leq 5$ components when decomposed in $Y_{lm}$ basis, which suffices our calculations. Specifically, for the 12-point icosahedron integral quadrature, we approximate the integration 
\begin{equation}
    J = \frac{1}{4\pi} \int d\bom~f(\bom) ,
\end{equation}
on a unit sphere by the sum 
\begin{equation}
\label{eq:quad}
    J = A \sum_{i=1}^2 f(a_i) + B\sum_{i=1}^{10} f(b_i) ,
\end{equation}
with the points $a_i$, $b_i$ written in spherical coordinates $(\theta,\phi)$ as:
\begin{equation}
    \begin{aligned}
    &a_1 = [0,0], a_2 = [\pi,0] \\
    &b_i = \begin{cases}
    \left[\arctan 2,  \frac{2\pi}{5} (i-1) \right] & \text{for } 1\leq i\leq 5 \\
    \left[\pi - \arctan 2,  \frac{\pi}{5}(2i-11) \right] & \text{for } 6\leq i\leq 10
    \end{cases}
    \end{aligned},
    \label{eq:ecp_details:quadrature_points}
\end{equation}
and the coefficients $A$, $B$ are 
\begin{equation}
    A = B = \frac{1}{12}.
\end{equation}
%
%
The coordinates of the $12$ points are written in a certain coordinate system, whose orientation remains arbitrary and each choice can be related by a global rotation. Generally speaking, for an arbitrary integrand, its integral result varies in a different coordinate system, since the employed 12-point icosahedron quadrature is only exact for $l\leq5$ component. In order to cancel out the error introduced by the orientation choices, we could average the results from randomly chosen orientations, then the result is actually an unbiased Monte Carlo estimation on a unit sphere. 

In our calculation, however, we only chose one random orientation for each sample at one Monte Carlo step
to have satisfactory training speed with some sacrifices in the variance of the energy estimation. Although, more orientation configurations in each estimation step would help decrease the variance, in our calculation, a one-orientation estimation has already given a satisfactory variance compared to that introduced from the MCMC steps.




\subsection{Calculation Specification}
Based on the open-sourced FermiNet package \cite{pfau2020ab,spencer2020better}, the numerical setup for our calculations is listed in Table~\ref{tab:exp_env}. Note that in most cases, we used the default values for hyper-parameters provided in the open-sourced repository for FermiNet.
The neural networks are trained on the internal machine learning platform in ByteDance Inc., which supports elastic resources and large-scale training tasks.

\begin{table}[ht]
\caption{\label{tab:exp_env} Numerical Setup}
\begin{ruledtabular}
\begin{tabular}{l|c} 
Name& Value \\
\hline
Framework & JAX \cite{jax2018github} \\
Main computing resource & 8 Nvidia V100 GPU cards \\
Optimizer & KFAC \cite{kfac} \\
Optimization hyper-parameters & Same as in Refs.~\cite{pfau2020ab, spencer2020better}\\
Batch size & 4096 \\
Number of HF pretrain iteration & 3000 \\
Number of training iteration & 500,000 \\
Number of inference iteration & 100,000 \\
Pretrain basis & ccECP-cc-pVDZ \cite{ecp_3d} \\
\end{tabular}
\end{ruledtabular}
\end{table}


\section{Results}
\label{sec:result}
 
 
 In this work, we mainly study 3d transition metals, namely from Sc to Zn, and their monoxides. 
 Their ground state energy and dissociation energy (DE) are relatively difficult to compute using traditional methods\footnote{Results calculated with various methods can be found in Refs.~\cite{ecp_3d,ecp_al,jiang2021full,PhysRevX.10.011041}.}.
 Moreover, ccECP can remove a significant amount of core electrons from calculations for those systems. For instance, atom Sc has 21 electrons in total, while we only need to consider 11 electrons with ccECP. We also performed calculations on elements Ga and Kr, where most electrons are treated as core electrons by ccECP, so as to show benefits on computation efficiency. From now on, we refer to our ECP based FermiNet method as FermiNet + ECP.
 

For ground state energy of 3d transition metals, our results are close to the state-of-the-art CCSDT(Q) at CBS limit and outperform projection methods such as fixed-node diffusion Monte Carlo (DMC)~\cite{ecp_al}. For dissociation energy of transition metal monoxides, our results are consistent with highly accurate CCSD(T), semistochastic heat bath configuration interaction (SHCI), auxiliary field quantum Monte Carlo (AFQMC) and density matrix renormalization group (DMRG) results~\cite{ecp_3d,PhysRevX.10.011041}.

\subsection{Atoms}
In this section, we study the ground state energy of atoms using FermiNet with ccECP. We carry out calculations on elements Ga, Kr and 3d transition metals, with the results listed in Table \ref{tab:ecp_energy_on_atoms}. For the purpose of comparison, we also list CCSDT(Q) and DMC results with ccECP from Ref.~\cite{ecp_al}. Note that all the compared methods are dealing with the same effective core Hamiltonian since they use the same ECP, hence it is fair to make comparisons on the ground state energy. 
The lower the energy, the better the method is.

\begin{table*}[t]
\caption{\label{tab:ecp_energy_on_atoms} Calculated ground state energy in Hartree for Ga, Kr and 3d transition metals, using various methods with ccECP: FermiNet + ECP, CCSDT(Q) with both cc-pCV5Z basis and CBS limit, and DMC using millions of determinants in its trial wave function. The results of the latter three methods are from Ref.~\cite{ecp_al}. CCSDT(Q) at CBS limit achieves the lowest energy for all the systems, and our method performs the second best. For heavier transition metals such as Cu and Zn, our calculated energies are around 10 mHa higher than the CCSDT(Q) / CBS ones, while the other methods' results are more than 40 mHa higher.}
\begin{ruledtabular}
\begin{tabular}{c|cccc} 
    \thead{Element} & \thead{FermiNet + ECP} & \thead{CCSDT(Q) / cc-pCV5Z } &
    \thead{CCSDT(Q) / CBS } & \thead{DMC(MD)} \\ 
    \hline
    Ga  & -2.039853(2) & -2.0395 & -2.039915(13) & -2.0392(2)  \\ 
    Kr & -18.472355(8) & -18.4659 &  -18.47259(27) & -18.4680(1) \\
    \hline
    Sc  & -46.5536(1) & -46.5494 & -46.55704(81) & -46.550(1)\\
Ti & -58.0895(1) & -58.0826 &-58.09263(76) & -58.085(1)\\
V  & -71.4346(1) & -71.4274 & -71.44178(59) & -71.421(2)\\
Cr & -86.6349(1) & -86.6236 & -86.64109(33) & -86.625\\
Mn & -103.8828(10) & -103.8718 & -103.8919(10) & -103.859(2)\\
Fe & -123.3754(1) & -123.3626 & -123.38804(93) & -123.358(2)\\
Co & -145.1409(1) & -145.1232 & -145.1541(10) & -145.115(4)\\
Ni & -169.3760(1) & -169.3546 & -169.3912(12) & -169.345(2)\\
Cu & -196.3873(1) & -196.3584 & -196.4038(10) & -196.353(3)\\
Zn & -226.3516(2) & -226.3210 & -226.3699(18) & -226.320(3)\\
\end{tabular}
\end{ruledtabular}
\end{table*}

\begin{figure}[htb]
\centering
\includegraphics[width=0.48\textwidth]{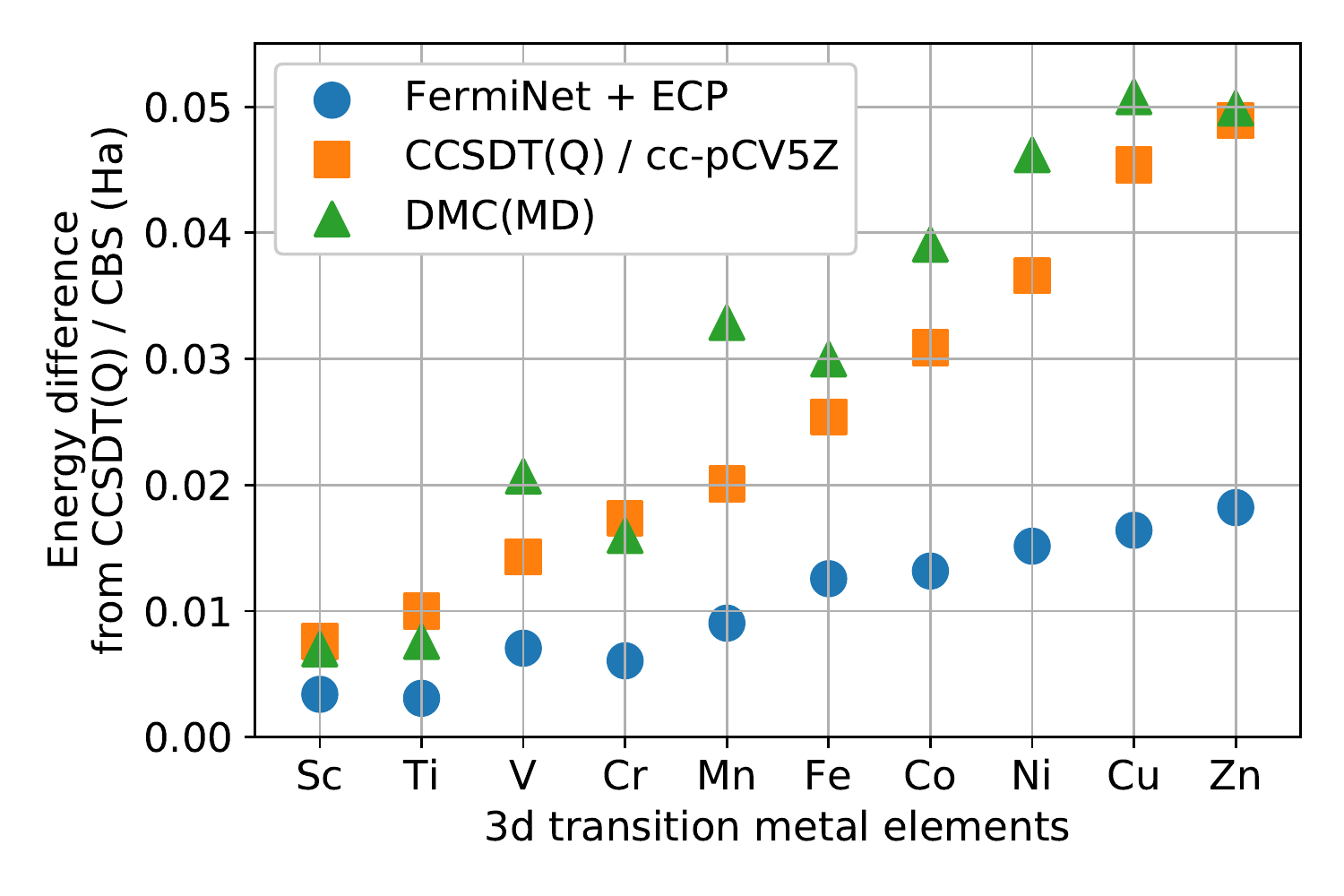}
\caption{\label{fig:metal_ecp_energy} Ground state energy of 3d transition metals calculated using FermiNet + ECP (blue dots), compared with CCSDT(Q) / cc-pCV5Z (orange squares) and DMC (green triangles) results provided in Ref.~\cite{ecp_al}, all with ccECP. Here we show the relative difference against CCSDT(Q) / CBS. It is clear that our method achieves better results than the compared ones, even though they are still around 10 mHa higher than the referential result for heavier atoms.}
\end{figure}

Among all compared methods, CCSDT(Q) at CBS limit, despite its non-variational nature and basis-set extrapolation error, has the lowest energy for all considered systems, and thus we use its results as the reference, against which the difference from other results is plotted in Figure~\ref{fig:metal_ecp_energy}. Our method achieves the second-best result. In particular, our result is better than the CCSDT(Q) with basis cc-pCV5Z, the largest finite basis set used in CCSDT(Q) calculation for those atoms to our knowledge, which is as expected since neural-network related ansatz has the ability to approach CBS limit~\cite{doi:10.1063/5.0032836}. Fixed-node DMC result with multi-reference trial wave function is also plot in Figure 1. DMC typically requires millions of determinants in its trial wave function in order to decrease the fixed-node error. In comparison, neural-network based methods can achieve better accuracy than DMC method with only dozens of determinants. Moreover, the gap between results from CCSDT(Q) / cc-pCV5Z and CBS limit increases quite significantly as the number of electrons increases, and similarly for DMC. Comparatively, the discrepancy from our results to the referential ones is much smaller for heavier transition metals such as Cu and Zn, suggesting that ECP method works well with FermiNet in those larger systems.

\subsection{Transition Metal Monoxides}\label{ssec:tmm}
In order to show our methods can be used to obtain high quality \textit{ab initio} data comparable to experiments, we compute
the dissociation energy ${\rm DE}$ of 3d transition metal monoxides, defined as 
\begin{equation}
\label{eq:de}
    {\rm DE}({\rm X})={\rm E}({\rm X})+{\rm E}({\rm O})-{\rm E}({\rm XO}),
\end{equation}
where ${\rm XO}$ denotes the monoxide of element X and ${\rm E(\cdot)}$ denotes the ground state energy. For monoxides, we only apply ECP to transition metal atoms while all electrons for atom O are included, and we use $-75.06655\ {\rm Ha}$ for ${\rm E}({\rm O})$ (same as the result given in the original FermiNet paper \cite{pfau2020ab}). 
 
\begin{figure}[h]
\centering
\includegraphics[width=0.48\textwidth]{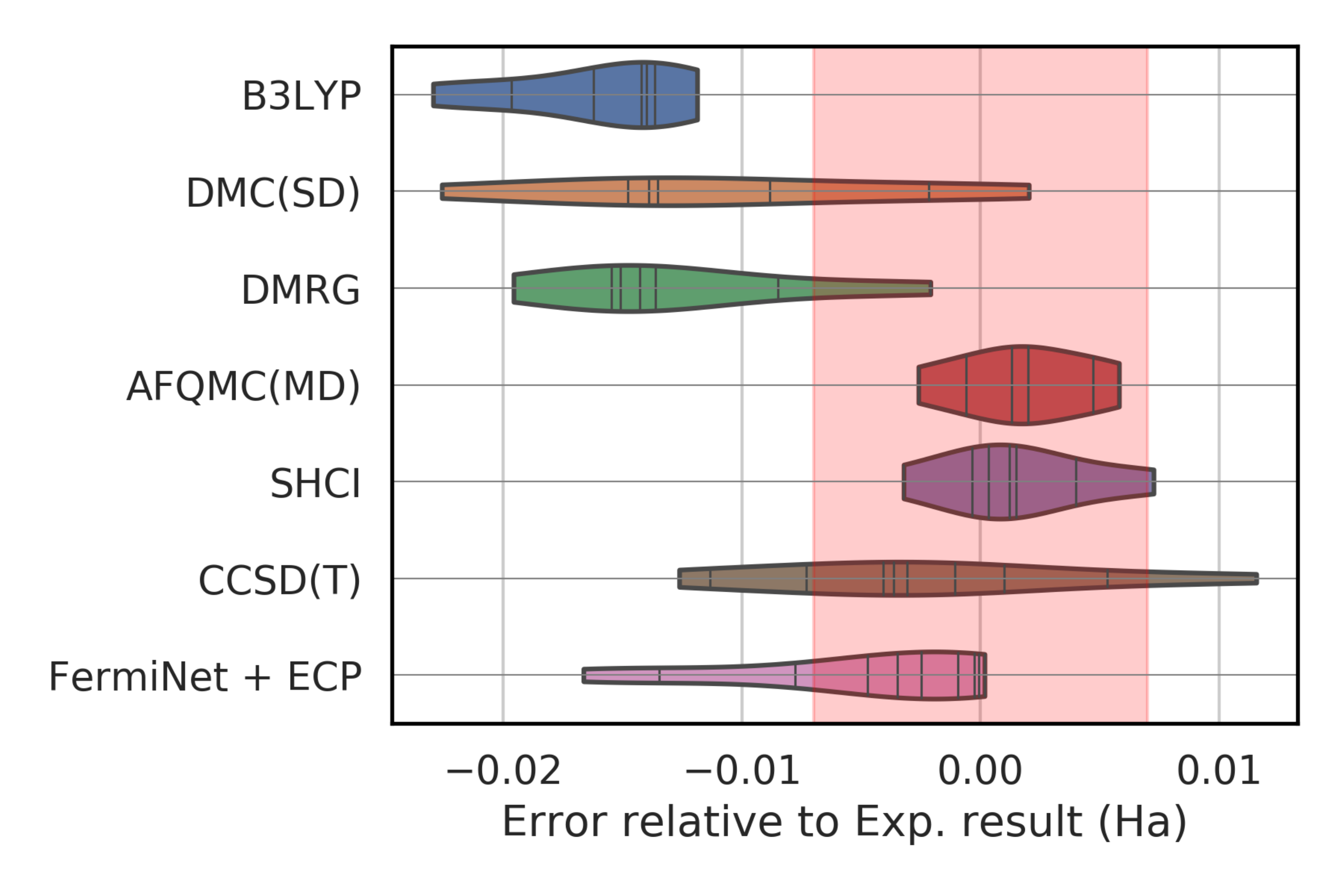}
\caption{\label{fig:bde_violin.eps} Fitted density of the difference from calculated transition metal monoxide DE to the experimental data~\cite{exp_data} in Hartree for each method. The CCSD(T) data (with ccECP) is from Ref.~\cite{ecp_3d} while others (with TN ECP \cite{trail}) are from Ref.~\cite{PhysRevX.10.011041}. Individual data points are indicated by small vertical lines. The red shaded region, centered at 0 and with width 0.007 Hartree, indicates the acceptable discrepancy from calculated results to experimental data. All the data from SHCI and AFQMC, as well as most data from our method and CCSD(T) fall into that region.}
\end{figure}

\begin{table*}[htb]
\caption{\label{tab:monoxide} Comparison of calculated DE of 3d transition metal monoxides against the experimental data \cite{exp_data}, the CCSD(T) / CBS result~\cite{ecp_3d}, and the calculated results from SHCI, AFQMC and DMRG~\cite{PhysRevX.10.011041}. CCSD(T) results use ccECP and the associated basis while SHCI, AFQMC and DMRG use TN ECP. Specifically, SHCI and AFQMC use aug-ccpV5Z basis and DMRG use aug-ccpVDZ basis. Our DE result is calculated with the ground state energy of both monoxides (listed here) and atoms (listed in Table~\ref{tab:ecp_energy_on_atoms}), and we use -75.06655 Ha as the ground state energy for atom O~\cite{pfau2020ab}. The bond length of monoxides in our calculation is the same as the one in CCSD(T) result~\cite{ecp_3d}.}
\begin{ruledtabular}
\begin{tabular}{c|c|cccccc} 
    \multirow{2}{*}{System} & \multirow{2}{*}{Ground State Energy (${\rm Ha}$)} &
    \multicolumn{6}{c}{Dissociation Energy (${\rm Ha}$)}  \\
    &&\thead{FermiNet + ECP} & \thead{Experimental Data} & \thead{CCSD(T)} & \thead{SHCI} & \thead{AFQMC} & \thead{DMRG}  \\ 
    \hline
    ScO &  -121.8765(1) & 0.2564(2) & 0.2566(3) & 0.2550(11) & 0.25684(1) & 0.2585(16) & 0.2422 \\ 
    TiO & -133.4073(1) & 0.2513(2) & 0.2548(25) & 0.2517(11) & 0.25444(1) & 0.2568(16) & 0.2397\\
    VO & -146.7417(2) & 0.2406(3) & 0.2405(31) & 0.2464(7) & 0.24439(1) & 0.2462(15) & 0.2268\\
    CrO & -161.8682(13) & 0.1668(14) & 0.1670(22) & 0.1684(3) & 0.17424(1) & 0.1717(22) & 0.1649 \\
    MnO & -179.0892(2) &  0.1399(12) & 0.1423(29) & 0.1390(7) & 0.14377(1) & 0.1436(18)  & 0.1268\\
    FeO &  -198.5927(1) & 0.1508(2) & 0.1555(3) & 0.1525(11) & 0.15671(1) & 0.1549(18) & 0.1359\\
    CoO & -220.3459(2) & 0.1385(3) & 0.1519(33) & 0.1396(18) & NA & NA  & NA \\
    NiO & -244.5855(4) & 0.1430(5) & 0.1439(11) & 0.1561(11) & NA  & NA  & NA \\
    CuO & -271.5504(16) & 0.0965(17) & 0.1131(11) & 0.1010(6) & 0.10987(1) & 0.1105(25) & 0.1046\\
    ZnO & -301.4705(3) & 0.0523(5) & 0.0608(14) & 0.0536(18) & NA &  NA & NA \\
\end{tabular}
\end{ruledtabular}
\end{table*}

We compare our calculated DE against high-accuracy electronic structure methods such as SHCI, CCSD(T), DMC, DMRG and AFQMC in Table~\ref{tab:monoxide}. 
Moreover, since DFT is broadly employed in computational studies of transition metal systems, we also list DFT results using the B3LYP exchange correlation functional for comparison. 
All these compared results are from Refs.~\cite{PhysRevX.10.011041,ecp_3d}, where we only take results calculated with the largest basis set provided. We use the same equilibrium bond length for monoxides as CCSD(T) results in the supplementary material of Ref.~\cite{ecp_3d}. 
For completeness, we also list our calculated ground state energy of monoxides in Table~\ref{tab:monoxide}. 
 
To visualize the overall performance of FermiNet + ECP calculation, in Figure~\ref{fig:bde_violin.eps} we plot the fitted density of the difference between the calculated result and the experimental data~\cite{exp_data} of 3d transition metal monoxide DE for each method. Here we compare with the experimental data because they can serve as reasonable references when comparing methods using different ECPs and basis sets. 
However, it is worth noting that experimental results are not completely self-consistent among different studies, which may differ from one to another by as much as 0.5 eV~\cite{PhysRevX.10.011041,exp_data}.
Therefore, we place a shade region (with a width of 0.007 Ha.) on Figure~\ref{fig:bde_violin.eps} to indicate acceptable discrepancy from the experimental data.
All data from very accurate methods such as AFQMC and SHCI fall into this interval. 
Our results mostly fall into this interval except for a long tail on the left, led by the CoO, CuO and ZnO results. 
We will comment on the issue associated with those monoxides later, but overall our calculations provide results almost as accurate as the current state-of-the-art electronic structure approaches.
DMRG, DMC and B3LYP data have less overlap with this interval.
We want to point out that those methods perform less ideally for different reasons: DMRG is known to be an accurate method, but the heavy cost limits its calculation of these molecules to aug-ccpVDZ basis set. So the basis set incompleteness error has dominated its DE underestimation. 
DMC is largely limited by the fixed-node approximation and the single determinant trial wave function employed.
For transition metal oxides, multi-configurational wave function is needed and fixed-node DMC based on single determinant trial wave function does not perform satisfactorily. 
For DFT, it is known that the results strongly depends on the chosen exchange correlation functional. 
The comparison of different functionals is beyond the scope of this work, but we present the results of one functional, namely the B3LYP functional, which is a hybrid functional known as one of the well-behaving functionals for transition metal oxides. 
Yet the results shows that B3LYP underestimates the DE by an average of 15 mHa.
Therefore, we conclude that FermiNet + ECP is more accurate and reliable than DFT, and should be promoted in future studies to tackle larger transition metal containing molecules and materials. 

\begin{figure}[h]
\centering
\includegraphics[width=0.48\textwidth]{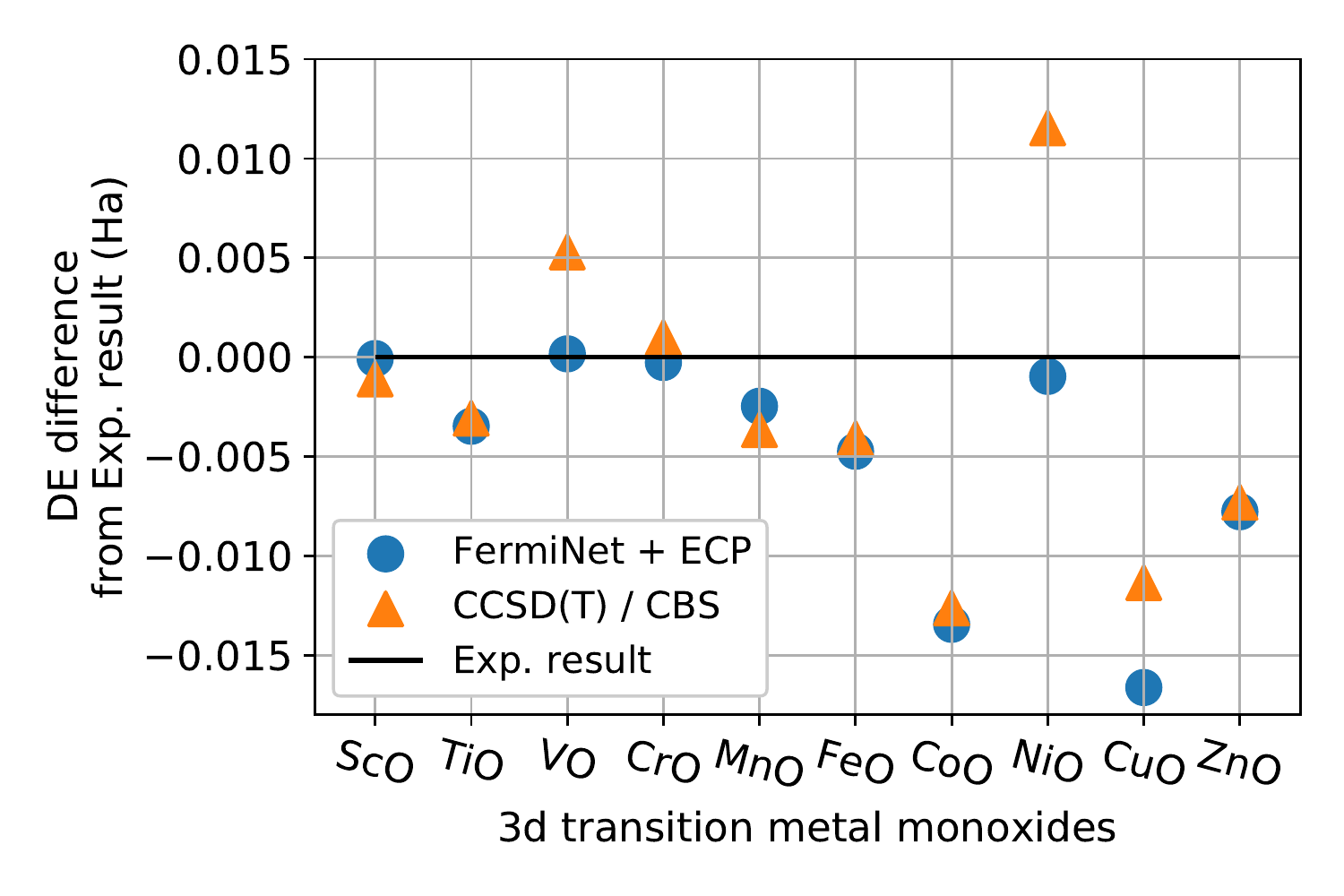}
\caption{\label{fig:bde} DE of transition metal monoxides calculated using FermiNet + ECP (blue dots) compared to CCSD(T) (orange triangles) both with ccECP. We show the relative difference against the experimental data. Our results are quite close to the CCSD(T) ones, except for VO, NiO and CuO, in which cases our results are closer to the experimental ones for VO and NiO.}
\end{figure}

The fitted density shape of the CCSD(T) result is similar to ours, but it has long tails on both sides. 
The good performance of CCSD(T) on transition metal monoxide was also mentioned in a recent study using full configuration interaction quantum Monte Carlo \cite{jiang2021full}, here we show that FermiNet + ECP outperforms CCSD(T) slightly for the majority of 3d transition metal monoxides.
In Figure~\ref{fig:bde} we present a finer comparison between our method and CCSD(T) at CBS limit~\cite{ecp_3d}, where the difference to the experimental data is compared for each system. 
For most of the monoxides, our results are close to the CCSD(T) ones, except for VO, NiO and CuO. For VO and NiO, our results are very close to the experimental data.  For CuO, both our result and CCSD(T) one are quite far away from the experimental reference (greater than 10 mHa), while ours are farther away. We will discuss more on our CoO, CuO and ZnO results in the next section. Note that compared to CCSD(T), known as the golden-standard in quantum chemistry, our method has not only comparable accuracy, but also better computational scaling, namely $N^4$, same as FermiNet, as opposed to CCSD(T)'s $N^7$, where $N$ stands for the number of electrons considered in the computation.

\section{discussion}
\label{sec:discus}
In this section, we discuss some details of our calculation and model training.

\subsection{Efficiency}
Efficiency and scalability, as well as accuracy, are our main motivations for taking ECP into consideration. 
ECP reduces the number of electrons to deal with in FermiNet, but it also introduces additional cost of numercial intergration of the ECP Hamiltonian.
In neural networks, the cost of such numerical interaction has not been investigated.
Whether the combination of FermiNet with ECP can have better computation efficiency and scalability than all-electron (AE) calculations needs to be examined.

Here we show the application of FermiNet + ECP on elements Ga and Kr in which cases ECP method has a significant advantage. With ccECP, we only need to consider 3 electrons for Ga and 8 electrons for Kr, while with all-electrons, we have to deal with 31 and 36 electrons for Ga and Kr respectively, which are already quite large systems for FermiNet to handle. The computation runtime are listed in Table~\ref{tab:ecp_runtime_on_atoms}, where the data is obtained from 1000 training iterations by taking the average of their runtime, and we also ignore the first several iterations to avoid the effect of initial warmup and/or compilation process. It is apparent that, compared with the AE calculation, the simulation efficiency is highly improved with ECP employed. 

In Table \ref{tab:ecp_runtime_on_atoms} we also list the computation runtime of 3d transition metal atoms, for which 10 core electrons are considered in ccECP. Because of the fixed number of core electrons, the heavier the element, the smaller the difference between AE and ECP is. 
For Sc there is about 50\% reduction in single-iteration runtime from ECP, whereas for Zn the reduction is less than 20\%. 
Nonetheless, we find that our implementation of ECP in FermiNet has larger gains from electron reduction than losses due to integration of ECP part, which suggests a promising future for applying various ECPs in fast-developing neural network electronic structure packages.

\begin{table}[t]
\caption{\label{tab:ecp_runtime_on_atoms} The number of electrons and runtime of a single training iteration in AE and ECP calculation respectively.}
\begin{ruledtabular}
\begin{tabular}{c|cc|cc} 
\multirow{2}{*}{Element} & \multicolumn{2}{c|}{AE} & \multicolumn{2}{c}{ECP} \\
& \thead{\#Electrons} &  \thead{Single iteration \\ runtime (s)} & \thead{\#Electrons} &\thead{Single iteration \\ runtime (s) } \\
    \hline
 Ga & 31 & 2.9 & 3 & 0.086\\
 Kr & 36 & 4.5 & 8 & 0.39\\
     \hline

Sc & 21 & 1.08 & 11 & 0.61\\
Ti & 22 & 1.19 & 12 & 0.71\\
V  & 23 & 1.32 & 13 & 0.85\\
Cr & 24 & 1.42 & 14 & 0.99\\
Mn & 25 & 1.65 & 15 & 1.13\\
Fe & 26 & 1.80 & 16 & 1.30\\
Co & 27 & 2.00 & 17 & 1.55\\
Ni & 28 & 2.11 & 18 & 1.73 \\
Cu & 29 & 2.32 & 19 & 1.99\\
Zn & 30 & 2.59 & 20 & 2.16 \\
\end{tabular}
\end{ruledtabular}
\end{table}


\subsection{Optimizer Comparison}
\begin{figure}[htb]
\centering
\includegraphics[width=0.50\textwidth]{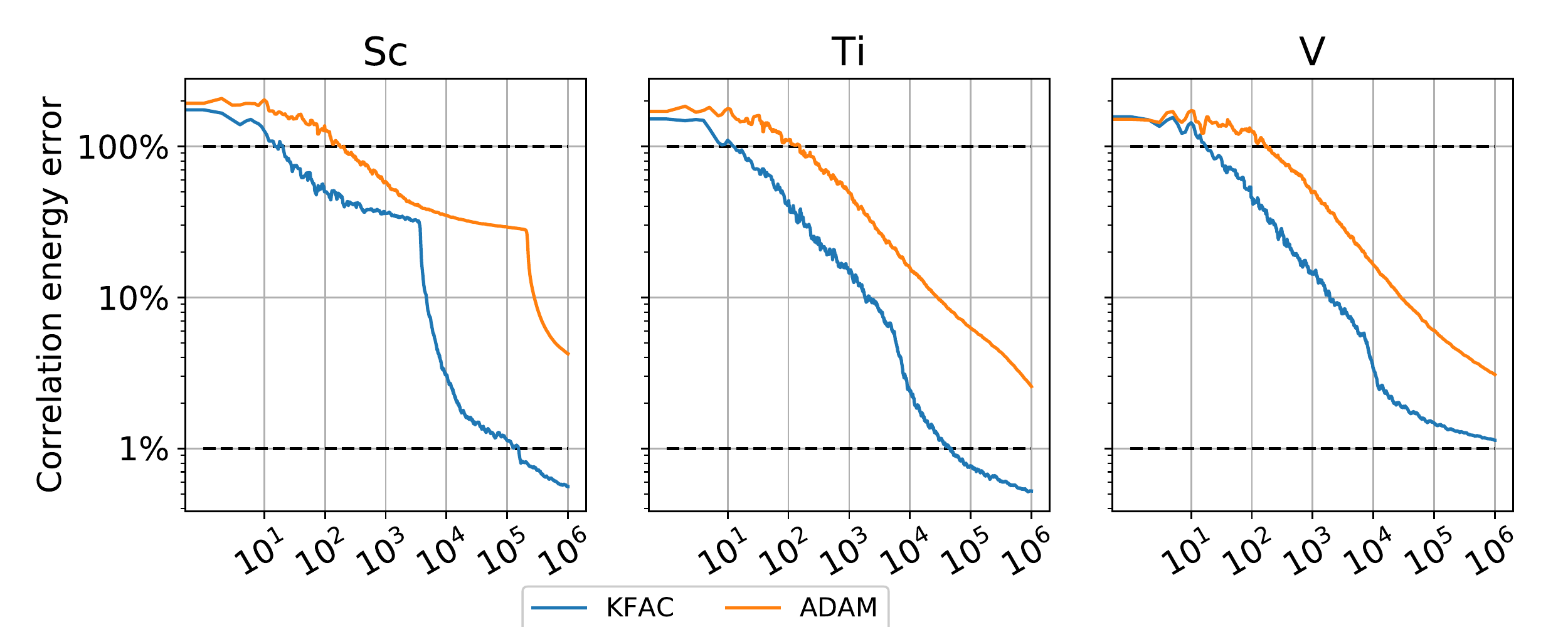}
\caption{\label{fig:adam_kfac.eps} The log-log plot of the optimization progress of Fermint with ECP for Sc, Ti and V using optimizers KFAC (blue lines) v.s. ADAM (orange lines). The horizontal axis is the number of training iterations. The vertical axis is the correlation energy error, calculated using CCSDT(Q) / CBS with ccECP provided in Ref.~\cite{ecp_al}. For clarity, we show the median energy over the last 10\% of iterations. }
\end{figure}

Following the original work of FermiNet \cite{pfau2020ab}, we have tested both ADAM \cite{DBLP:journals/corr/KingmaB14} and KFAC \cite{kfac} optimizers when training neural networks with ECP. In terms of training runtime, we find that the choice of optimizer does not matter much, suggesting that the runtime is dominated by the forward pass, especially the numerical integration introduced by ECP method. 
For instance, for the vanadium atom, one training step runs for around 0.8 seconds regardless of the optimizer. 
As for the model performance, KFAC can lead the neural network to a better state yielding lower energy. It also converges in much fewer iterations. As shown in Figure~\ref{fig:adam_kfac.eps} where we show atoms Sc, Ti and V as examples. 
Energy optimized by KFAC approaches 1-percent of correlation energy error after around ten thousand iterations while the one optimized by ADAM is not close to that level of accuracy even after one million iterations.

\subsection{Model Training Details}

Here we discuss a few technical issues occurred in calculating Co, Cu, Zn and their monoxides, which may be related to the bad performances presented in above sections. 
The first issue is that when training the neural network for these atoms, it may start to produce Not-a-number (NAN) value after a number of iterations, which ruins the whole process.
To resolve this issue, we removed outliers in terms of the local energy from the training process instead of simply doing clipping. 
Surprisingly, there was a magical threshold for outlier identification such that the training energy may be around 10 mHa lower than the result with a lower or higher threshold. 
In our case, such threshold is 10 times the standard deviation of the local energy in one training batch. 
We did not observe such phenomena for monoxides and we speculate that it is because electrons are less likely to be too close to the nuclei of transition metal atoms in molecules.
Note that during inference phase we do not do such outlier removal so that our calculated energy remains reliable and variational. 
Therefore, although this is an issue to be further investigated in future developments of FermiNet, we do not expect it is the main cause of the deviation of our results from experimental values.

Moreover, as mentioned in Section \ref{ssec:tmm}, our calculated DEs for CoO, CuO and ZnO are underestimating comparing to the result of the most accurate methods and the experimental data. CoO result is 13 mHa lower, CuO result is 17 mHa lower and ZnO result is 8 mHa lower, in which cases the gap is more than $10\%$ of their DE. 
Since our method is variational, it is more likely that there is an overestimation of the total energy of the monoxide. 
There are several potential reasons. Firstly, we used the same bond length as CCSD(T)~\cite{ecp_3d} for monoxides' equilibrium position, which may not be the optimal value for FermiNet + ECP. 
A fine search of the optimal bond length may improve the calculated DE by a few mHa. 
Secondly, those monoxides are notoriously strongly correlated and challenging systems for electronic structure calculations. 
Although we have made it possible with FermiNet + ECP, the system is at the limit of current computational capacity. 
Thus, it remains unclear whether a more powerful network with a larger number of layers and determinants can improve the description of such strongly correlated systems.
Last but not least, we trained our neural networks up to 500,000 iterations, up to which the convergence of the training process becomes really slow.
This maximum number of training iterations is chosen based on our affordability, but the slow convergence does indicate that the wave function represented by the neural network is still evolving slowly.
For such challenging systems, we can not rule out the possibility that further fine-tuning in the network can eventually lead to better results.
How the network size and training process affect the performance of neural networks is certainly an interesting topic for future studies, especially for strongly correlated systems.

\vspace{-0.5em}
\section{Conclusion}
\label{sec:conclu}
In this work, we have implemented the ECP method in the existing deep learning work, namely FermiNet. Employing ECP method pushes FermiNet towards studying larger systems by reducing the cost while retaining the accuracy to describe chemical bonding.
Based on our implementation we carry out studies of 3d transition metal atoms and their monoxides, which are challenging systems for electronic structure methods. The calculated results are consistent with state-of-the-art computational methods and experimental data.
Comparing to more widely used methods such as DFT and CCSD(T), FermiNet + ECP outperforms them in a broad range of systems, suggesting a bright future for deep learning techniques in molecules and materials modelling.
Based on this study, we expect ECPs can also be successfully used in other neural networks, where benefits in computational efficiency are likely to be similar to our work.
We also want to mention that 
investigating better network structures, such as the Transformer network structure \cite{NIPS2017_3f5ee243}, may further improve the accuracy and efficiency in solving electronic structures, and related works are under progress.


\vspace{1.0em}

\section*{Acknowledgements}

The authors thank Matthew Foulkes, David Ceperley and Abdulgani Annaberdiyev for helpful discussions, Hang Li for support and guidance, DeepMind FermiNet team for the open-sourced FermiNet software and prompt response on Github. Additional thanks to the ByteDance AML team for technical support, ByteDance AI-Lab LIT Group for fruitful collaboration, and the rest of ByteDance AI-Lab Research team and collaborators for ideas and inspiration.



\bibliography{reference}

\end{document}